# Inverse spillover and dimensionality effects on interstitial hydrogen

K. Komander*[1], G.K. Pálsson[1], S.A. Droulias[1], T. Tsakiris[1], D. Sörme[1], M. Wolff[1], D. Primetzhofer[1]

[1]*Department of Physics and Astronomy, Uppsala University, Box 516, S-751 20 Uppsala, Sweden*
*kristina.komander@physics.uu.se

Nanoscaling interstitial metal hydrides offers opportunities for hydrogenation applications by enhancing kinetics, increasing surface area, and allowing for tunable properties. The introduction of interfaces impacts hydrogen absorption properties and distribution heterogeneously, making it however challenging to examine the multiple concurrent mechanisms, especially at the atomic level. Here we demonstrate the effect of proximity on interstitial hydrogen in ultrathin single crystalline vanadium films, by comparing hydride formation in identically strained Fe/V- and Cr/V-superlattices. Pressure concentration and excess resistivity isotherms show higher absolute solubility of hydrogen, higher critical temperature and concentration in the Cr/V-superlattice. Direct measurements of hydrogen site location and thermal vibrations show identical occupation of octahedral z sites at room temperature with a vibrational amplitude of 0.20-0.25 Å over a wide range of hydrogen concentrations. Our findings are consistent with a more extended region of hydrogen depletion in the vicinity of Fe compared to Cr, which showcases an inverse of the hydrogen spillover effect. Advancing the understanding of interface effects resolves previously puzzling differences in the hydrogen loading of Fe/V- and Cr/V-superlattices and is relevant for advancing both catalysis and storage.

Hydrogen in metals can occupy interstitial sites, enabling high hydrogen storage capacity [1], yet simultaneously cause highly detrimental effects like hydrogen-induced embrittlement [2]. Exhibiting high hydrogen mobility even at lower temperatures, interstitial hydrides have been used for purification, hydrogen sensing, and act as catalysts [3]. Hydrogen absorption is strongly affected by composition and structure at the atomic scale, which provides opportunities to improve reaction kinetics and/or reduce the enthalpy of formation to meet requirements for applications [4]. Particularly, interfaces influence metal hydride formation through a number of mechanisms which are challenging to disentangle. For example, strain due to elastic boundary conditions at interfaces alters the local electronic structure at interstitial sites leading to changes in local binding energies of H and elastic H-H-interaction [5] which affects thermodynamic properties [6] [7], site occupation [8] [9], and thereby also diffusion [10]. In the case of magnesium, partial hydrogenation leads to the formation of localized $MgH_2$ clusters, where electronic structure changes induce interface polarization slowing hydrogen sorption kinetics and leaving surfaces partially hydrogen-free [11]. In contrast, the solubility at surface and subsurface absorption sites of Pd can be enhanced due to weakened metal bindings and surface strains from fewer neighboring atoms [12]. Furthermore, missing hydrogen neighbors at surfaces of ultrathin films can decrease both the mean H-H-interaction and critical temperatures via finite-size effects [13] [14]. Effects of the surrounding chemical environment on the hydrogen-absorbing material, also called proximity effects, are, therefore, potential levers for tuning hydrogen uptake.

Interface effects can be studied by sandwiching different materials in superlattices with high crystal quality and identical strain states. Strain states of hydrogen-absorbing films with distinct adjacent non-absorbing materials can be tuned by layer thickness ratios and substrate adhesion [15] and are known to influence thermodynamic properties. Hence it is crucial to keep the initial strain state the same when investigating proximity. The hydrogen-absorbing material chosen in this study is vanadium, which is representative of a wide class of interstitial hydrides, absorbs hydrogen into distinct interstitial sites, and exhibits a rich thermodynamic phase diagram [16]. In bulk vanadium, hydrogen occupies different interstitial sites above and below the critical temperatures, T-sites (α-phases) and O-sites (β-phases), respectively and exhibits regions of co-existence and a suppressed spinodal [17]. The components neighboring V are either Fe or Cr, for which hydrogen absorption is endothermic at the studied conditions [18].

Previous attempts, focusing on reciprocal space techniques and density functional calculations, observed variations in hydrogen-induced lattice expansion and thermodynamic properties [19] [20] [21], which were attributed to site occupancy differences. Through careful re-analysis in this work, we reveal an underlying effect whose origin can be traced to the extension of hydrogen-depleted layers at the interfaces, explaining the observed differences in expansion coefficients without changes in site occupation. Consequently, an intrinsic finite-size effect is induced in the central layer due to a lowering of the effective thickness of the vanadium layer. While hydrogen spillover effects at interfaces [22] [23] [24] - the migration from one active site to another - have been shown to enhance hydrogen storage in a variety of materials, our findings suggest a contrasting phenomenon. Specifically, the absence of hydrogen near an interface with a non-absorbing metal, influenced by the type of adjacent element, corresponds to an inverse of hydrogen spillover. To investigate these effects, we employed direct real-space methods for determining hydrogen concentration, site location, and vibrational motion using energy-resolved and channeling nuclear reaction analysis (NRA) with $^{15}N$-ions [25] [26]. Phase boundaries were extracted via pressure composition and resistivity isotherms [27] using calibrated optical transmission as a proxy for hydrogen concentration. The origin of the optical response upon hydrogenation of Fe(Cr)/V-superlattices was investigated in [20] [28] [29] [30], and demonstrated to be dominated by lattice expansion.

**Thermodynamic analysis of confined hydride thin films in superlattices**

The inset in Fig. 1 illustrates the Fe/V- and Cr/V- superlattice samples with a thickness-ratio of 2/14 monolayers (ML) [15] on MgO substrates and covered with 7 nm of Pd to catalyze the dissociation of molecular hydrogen and to protect the sample from oxidation. Both superlattices have a crystal coherency comparable to the thickness of the films and are under close to similar tensile and compressive bi-axial strain in the plane with interlayer roughness of ±1 ML [19] [20]. The absorption of hydrogen is fully reversible and results in a one-dimensional out-of-plane volume expansion, while the samples remain metallic [7].

The pressure concentration isotherms (p-c-T) for Fe/V and Cr/V in Fig. 1 (a) reveal that the magnitude of hydrogen uptake at equal pressures scales significantly differently for the two samples. V in proximity to Cr absorbs approximately 1.5 times as much hydrogen compared to Fe, and is much closer to the bulk solubility [16]. Both exhibit onsets of pressure plateaus with a clear temperature dependence, while the plateau slopes and the equilibrium pressures are lower for Cr/V than Fe/V at the same temperature. In equilibrium, hydrogen pressure can be related to the chemical potential of the metal-hydrogen system, from which thermodynamic properties can be determined [31]. In the center of the plateau region, the derivative of the chemical potential approaches zero as the temperature is lowered, indicating the onset of a critical point in the region 0.05 H/V and 0.15 H/V for the two samples.

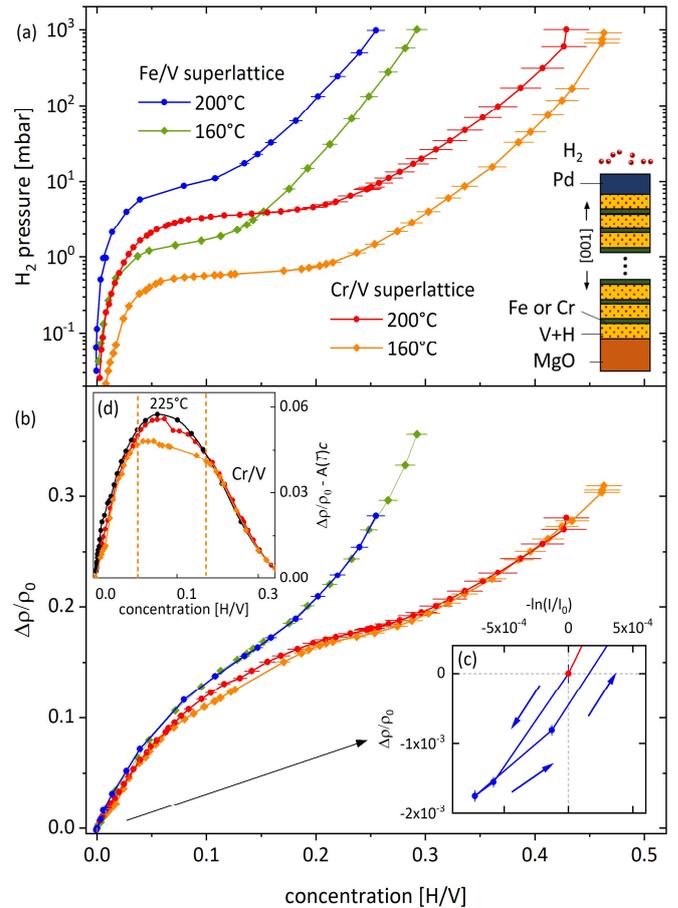

**Figure 1: Thermodynamic analysis of vanadium hydride films embedded in superlattices.**
(a) pressure-composition isotherms of Fe/V- and Cr/V- superlattices at 200°C and 160°C, hydrogen concentration refers to average hydrogen concentration in vanadium films extracted from calibrated optical transmission at 625 nm; (b) isotherms of change in excess resistivity $\Delta\rho$ normalized by $\rho_0$ without hydrogen; (c) $\Delta\rho/\rho_0$ enhanced for UHV to low-pressure region highlighting an initial decrease for Fe/V; (d) $\Delta\rho/\rho_0$ of Cr/V (225°C, 200°C, 160°C) subtracted by a linear function with a slope $A(T)$ ranging between 0.878-0.863 to enhance discontinuity from temperature-independent behaviour. The dashed lines mark the discontinuity edges at 160°C.

Phase boundaries are quantitively explored analyzing relative changes in the resistivity $\rho$ that result from scattering of the electrons by perturbations in the average potential landscape, and are sensitive to changes of hydrogen ordering in metals [32]. Fig. 1 (b) shows isotherms of excess resistivity, quantified by $(\rho-\rho_0)/\rho_0 = \Delta\rho/\rho_0$, where $\rho_0$ is the resistivity of the as-grown sample at temperature $T$. The $\Delta\rho/\rho_0$-c-$T$ isotherms of both superlattices are of similar shape, but resistivity changes in Cr/V extent further in average hydrogen concentration, similar to p-c-T isotherms in Fig. 1 (a). A minor initial decrease of $\Delta\rho$ in Fe/V for very low hydrogen pressures, visualized in Fig. 1 (c), indicates enhancement of crystal quality. This finding implies the presence of a low concentration of defects in the vanadium, such as vacancies, occupied prior to the interstitial sites, which cause the small concentration offset (2.5%-3.5%) in the calibration, as suggested in [30]. With increasing hydrogen content, the $\Delta\rho/\rho_0$-c-$T$ isotherms are close to independent of temperature; however, within a small region around ~0.1 H/V in Cr/V sudden drops in $\Delta\rho/\rho_0(c)$ appear below 200°C, which is demonstrated in Fig. 1 (d), enhanced by subtraction of a linear function from the data.

To analyze intrinsic changes in the electrical properties of thin metal hydride films, $\Delta\rho/\rho_0(c)$ has previously been expressed as a combination of several contributions [7] [33]. We separate the different features of $\Delta\rho/\rho_0(c)$ and correlate them to properties of the metal hydrides as follows: Bulk-like metal hydrides are known to exhibit a parabolic behavior in $\Delta\rho/\rho_0(c)$ for a continuous phase transition in the absence of phase separation [32], which we attribute to a term $\rho_H(c)$. The effect originates from conduction electron scattering from changes to the average potential. This scattering increases until the interstitial sites are half-occupied; beyond this point, the electrons scatter from the potential variations caused by the remaining unfilled sites. This behavior is seen in, for example, the α to α' transition in bulk V-H [34]. In Fig. 1 (b) and (d), such a parabolic behavior is observed centered around ~0.1 H/V, while superimposed by a second monotonically increasing contribution of a larger increase in Fe/V versus Cr/V. The effect has been observed in thin films and superlattices and is suspected to originate due to the presence of interfaces and hydrogen-depleted layers [33], and identified as a second contribution $\rho_{layer}(c)$. Deviations from the temperature-independent behavior, the sudden decreases in $\Delta\rho/\rho_0(c)$ around 0.1 H/V, highlighted in Fig. 1(d) for Cr/V, are associated with changes in $\rho_H(c)$. As variations in the ordering of lattice constituents influence the electronic scattering [35], the sudden changes in slope with decreasing temperature indicate a prompt change in concentration fluctuations. Similarly, sharp resistivity minima in $\rho_H(c)$ below the critical temperature are observed for the Ta-H system [32].

The hydrogen concentrations and temperatures corresponding to the discontinuity edges in $\rho_H(c)$, extracted by comparing $\Delta\rho/\rho_0(c)$ curves at various temperatures and identifying sudden changes in slope, are shown in Fig. 2 (solid circles). The concentration change to the nearest measurement yields a measure for uncertainty. We identify a phase boundary and approximate a parabolic shape, yielding a lower critical temperature and critical concentration in Fe/V, namely ~0.078 H/V at ~160°C versus ~0.133 H/V at ~196°C in Cr/V. The onset in Cr/V can already be observed at 200°C in Fig.1 (d). A lower critical temperature in Fe/V is also reflected in $p$-$c$-$T$ pressure plateaus of a lower incline than Cr/V at the same temperature. Plateaus are expected to be flat below the critical temperature due to equilibrium conditions between different phases. Sloped plateaus indicate a variation in stress distribution in the V layers.

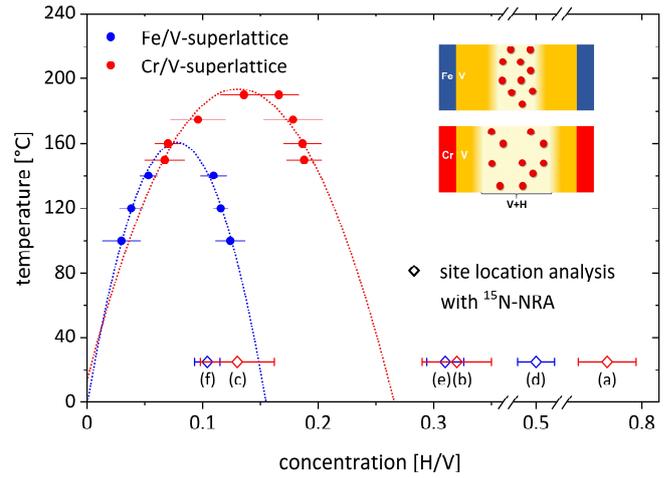

**Figure 2: Proximity effect on phase transitions of vanadium hydride.**
(solid circles) Phase boundaries for Cr/V- and Fe/V-superlattices obtained from sudden changes in slopes of $\Delta\rho/\rho_0$-c-T isotherms as indicated in Fig. 1 (d). Dotted lines follow a fitted parabolic function as a guide to the eye. (◊) mark hydrogen concentrations in the Cr/V (a-d) and Fe/V (d-f) where site location analysis is performed by channelling [15]N-NRA, presented in Fig. 3 (a-f). (inset) Illustration demonstrating hydrogen-depleted layers of different thicknesses lowering the effective thickness of the vanadium layer, which induces finite-size effects on the phase boundaries.

A decreasing ordering temperature has been associated with decreasing layer thickness arising from the missing H-neighbors at interfaces, similar to finite-size effects in magnetic layers and nanosized ice crystals [13] [36]. While the thickness of the vanadium films is identical for Fe/V and Cr/V, the thickness of the hydrogen-containing vanadium layer may not necessarily be the same if proximity to Fe induces a larger hydrogen-depleted region compared to Cr, as illustrated in the inset of Fig. 2. The nominal hydrogen concentration, normalized to the atomic fraction of V atoms in the superlattices [H/V], is averaged over absorbing and non-absorbing V layers. A variation in the amount of non-absorbing V for a fixed number of H atoms in absorbing V, results in different averaged hydrogen concentrations. Thus, the significant differences in average hydrogen uptake and critical concentrations support a proximity effect on the extent of the depletion layers. Hydrogen-depleted V layers have been reported at the interface to vacuum in [37] and to Fe. Extension of 4-5 Å have been visualized by atom probe tomography [38] and deduced from [15]N-NRA experiments [7], 2-3 Å (1.3-2) ML inferred from critical temperatures in [13] and 1 Å determined with neutron reflectometry [39].

Findings by W. Huang *et al.* [21] indicate a larger estimated depletion of 2.25 Å (1.4 ML) in Fe/V as compared to 1.6 Å (1 ML) in Cr/V.

Another observation from $p$-$c$-$T$ and $\Delta\rho/\rho_0$-$c$-$T$ isotherms in Fig. 1 is the reversible hydrogen ab- and desorption without the occurrence of hysteresis. Hysteresis is linked to coherent elastically driven phase transformations and interstitial sites changes [40] [41]. The site preference depends on the initial lattice strain and hydrogen-induced increase in strain, which scales differently depending on the type of site occupation. Out-of-plane lattice expansion coefficients of clamped vanadium were found to be 37%-50% larger for $O_z$- than for $T_z$-occupation due to the anisotropy of the local strain field [8] [42]. The concentration at which a site change occurs depends on whether hydrogen is ab- or desorbed, causing hysteresis in isotherm measurements. The absence of hysteresis in the vicinity of the phase boundaries thus suggests transitions without a change in hydrogen site occupancy. However, with increasing initial biaxial compressive strain of vanadium, the hysteresis gap is expected to decrease [8] and can, therefore, be very small in the fully strained superlattices. Further insights into site occupation can be gained, considering that the optical response, used to determine the hydrogen concentration for isotherms in Fig. 1, correlates with hydrogen-induced volume change [20] [30]. The isotherms obtained for $T<T_c$ and $T>T_c$, exhibit excellent overlap for $\Delta\rho/\rho_0$ and consistent changes in $p$-$c$-$T$ isotherms over the whole temperature range. These findings and the absence of hysteresis suggest that the observed phase transitions occur without a change in site occupancy, in stark contrast to the behavior of hydrogen in bulk vanadium. These results indicate that control over boundary effects can influence site occupancy and thereby inhibit deleterious changes, such as embrittlement, due to site changes and large differences in specific volumes of hydride phases.

**Real-space lattice location**

To test this hypothesis, site locations of interstitial hydrogen were directly investigated via $^{15}$N-NRA [26] at room temperature for different concentrations, indicated as (◊) in Fig. 2 for Cr/V (a)-(c) and Fe/V (d)-(f). The experiments (a), (b), and (d), (e) were conducted for average concentrations well above the phase boundaries, while (c) and (f), were performed within the parabolic phase boundary. The cross-sectional distribution of hydrogen atoms in superlattices was found to be homogenous, within the depth resolution of several nm. However, at the lowest concentrations, variations in the distribution were observed, as demonstrated in [30]. Site occupation is probed exclusively in the center of the hydrogenated Fe(Cr)/V superstructures and is derived from the resonant nuclear reaction yield with hydrogen (γ-yield) in channeling geometries, which directly correlates to the nuclear encounter probability (NEP) of $^{15}$N-ions to hydrogen interstitials. Fig. 3 shows the normalized channeling γ-yield for angular-resolved scans over the (110), (310), and (110) planes of the superlattice crystal. The observed channeling dips in the γ-yield relate to shadowing of hydrogen atoms by lattice atoms and demonstrate $O_z$ occupancy upon comparison to simulations of various site positions in [26]. Calculated NEPs of channelling $^{15}$N-ions to hydrogen in $O_z$ sites of different thermal vibrational amplitudes are plotted as solid lines. The corresponding residuals of the simulations to the experimental values, shown below in Fig. 3, can be used to judge the best fit. In Cr/V (a), (b) and Fe/V (d), (e), site occupation is identical; of $O_z$-type with thermal vibrational amplitudes of 0.20-0.25 Å. The observation of identical hydrogen vibrational amplitudes in both superlattices indicates similar vibrational characteristics and elastic properties of the metal hydrides within the measurement uncertainty. These thermal vibrations correspond to the zero-point motion of hydrogen in the potential well of the interstitial site, which is seemingly independent of the adjacent metal within the hydrogenated layer and the hydrogen concentration above the phase transition. Within the phase boundary, the γ-yield dips are less pronounced, more so in Fe/V (f) than in Cr/V (c), and $O_z$-occupation with 0.25 Å vibrational amplitude cannot reproduce the experiment within a 3σ-interval. The difference in the observed channelling pattern can result from occupation of both $T_z$ and $O_z$ sites, a larger vibrational amplitude or small displacements decreasing the shadowing of hydrogen atoms, or partial occupation of less shadowed or randomly distributed defect sites. For Fe/V in experiment (c), defect sites could account for a significant part of the hydrogen inventory [30], while the exact distribution of defect sites in the superlattice is unknown. Although the different configurations are, in principle, distinguishable by evaluating multiple geometries [26], due to large uncertainties of the normalized γ-yield, the quantitative analysis does not yield unambiguous results. The large uncertainty in γ-yield at low concentrations stems from a low signal-to-background ratio when employing low ion beam current to avoid beam-induced hydrogen loss. Consequently, the determination of the exact site locations from the data acquired in experiments (c) and (d), and also below the critical concentration, remains elusive under these conditions.

The observed site occupancy aligns with the theoretically predicted influence of axial compressive strain in [8], revealing the preference of $O_z$-sites for lattice strain above $c/a > 1.04$, which is reached at about 0.2 H/V in Fe/V, initially strained to $c/a = 1.01$ [26]. In confined thin films, $O_z$ occupation does not necessarily imply an ordered β-phase as in bulk vanadium, and a phase transition upon increasing the temperature beyond the critical temperature must not implicate a change in site occupancy, as suggested by the isotherm measurements in Fig. 1.

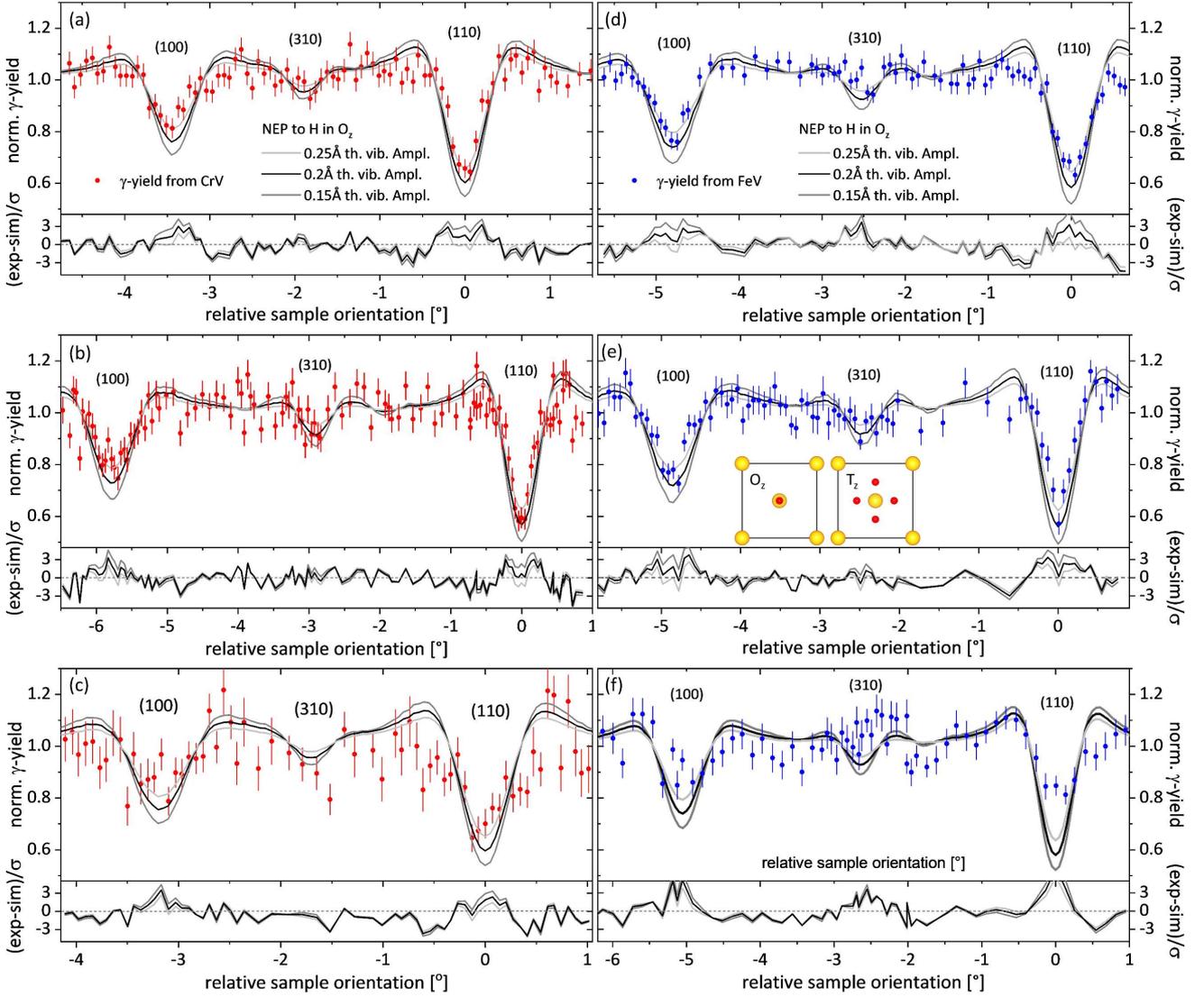

**Figure 3: Lattice site location and thermal vibration of interstitial hydrogen in vanadium.**
Normalized γ-channelling yield resulting from angular-resolved $^{15}$N-NRA for detection of hydrogen location in Cr/V (a)-(c) and Fe/V (d)-(f), loaded to different concentrations marked in Fig. 2 as (◊). The angular line scan crosses several crystal planes (100), (310) and (110) close to the [001] axis of the superlattices (bct). Hydrogen is probed exclusively in the centre of the superlattices corresponding to $^{15}$N-ion energies of 6.57 MeV and 6.55 MeV, respectively. The solid lines represent simulations of $^{15}$N-NEP to hydrogen in $O_z$ interstitial sites of various thermal vibrational amplitudes. The lower plots show the corresponding residuals normalized by the standard deviation $\sigma$. The insets in (e) illustrate projections of the vanadium (yellow) unit cell along [001] for hydrogen (red) in $O_z$- and $T_z$-sites.

**Proximity effects influence hydrogen distribution**

We have studied hydrogen absorption in single crystalline superlattices featuring stacks of ultrathin vanadium films in the proximity of Fe or Cr. Despite their identical structure and strain state, the average solubility of hydrogen is larger in vanadium in proximity to Cr, and phase boundaries are shifted such that the critical temperature and concentration are higher compared to vanadium layers in proximity to Fe. Furthermore, real-space site location below the transition temperature reveals identical site occupation of $O_z$-type with consistent thermal vibrational amplitudes, indicating indifference between the average zero-point energies. Optical transmission and excess resistivity measurements indicate that the site preference is temperature-independent within the range of 100-225°C.

These findings provide strong evidence of a larger hydrogen-depleted layer at the interface to Fe than to Cr, confirming earlier conjectures in [21], which indirectly yields a thinner absorbing layer, which in turn is subject to finite-size induced lowering of the critical temperature. Previous experimental studies on hydrogenated Fe(Cr)/V found enthalpy differences via optical transmission measurements [19]. Furthermore, deviating volume expansion above the phase transition temperatures of deuterated samples was observed via neutron reflectometry [20]. These findings can all be reconciled by the presence of differently-sized depletion layers at interfaces in V, an effect that displays an inverse of hydrogen spillover.

The origin of the depleted layers has been discussed in the literature. As alloying of V with Fe or Cr decreases hydrogen solubility [43] [44] [45], chemical intermixing at the interface contributes to hydrogen depletion at the interface. While both Fe and Cr have low negative enthalpies of mixing with V (-7 and -2 meV/atom), Cr has a slightly less negative enthalpy of mixing, thereby providing a lower driving force towards intermixing. Nevertheless, hydrogen-depleted regions in pure vanadium were separated from intermixed Fe/V by R. Gemma *et al.* [38] and were also observed near the vanadium surface to vacuum by E. Billeter *et al.* [37]. Modulated rigidity is considered in [46], as hydrogen distribution is governed by the energy cost of accompanying volume expansion, calculated to have a screening length of 2-3 ML. However, the shear modulus of Cr (115 GPa) is higher than Fe (82 GPa), implying the opposite effect observed in this work. Earlier studies suggested electrostatic repulsion of hydrogen due to charge transfer introducing interface dipoles [47] [48] [49]. While electrostatic effects are conceivable as Fe has two additional conduction electrons when compared to Cr, a theoretical study in [46] finds them mostly screened as electrons contribute to a common electron gas. Calculated densities of states presented in [19] demonstrate distinct impacts on V p- and d-bands at the interface with Cr and Fe, characterized by a small charge transfer from the first atomic V layer to Cr (0.2 e$^-$) and twice as large to Fe (0.4 e$^-$). The chemical binding of hydrogen around -7 eV was found unchanged both at the interface and in the center of the film. While the direct correlation to the electronic influences is not evident from these considerations, our findings suggest their potential role in the observed proximity effects on the hydrogen-depleted layer. Additionally, spin polarization resulting from magnetic effects is conceivable given that Fe is ferromagnetic and Cr antiferromagnetic; these, however, are to date unexplored.

The inverse spillover effect discussed here has large bearing on the fundamental understanding of the important spillover effect seen in other systems, and a careful, systematic approach at the atomic level holds considerable promise for elucidating its underlying mechanism. Once fully understood, spillover effects can be leveraged to tailor thermodynamic, kinetic and catalytic properties of metal hydrides, enhancing their performance in various applications.

**Experimental Methods**

*Superlattice samples*

The Fe/V- and Cr/V-superlattices are epitaxially grown on polished single-crystalline MgO(001) via dc magnetron sputtering [15]. The Fe(Cr)/V layers have a nominal thickness-ratio of 2/14 monolayers (ML), which minimizes the elastic energy in the bi-layer [15], and repeated 20 and 23 times, respectively. Elemental composition analysis employing Rutherford backscattering spectrometry (RBS) with $^4$He(2MeV)-ions reveals slightly higher Fe and Cr concentrations with a true ratio of about 1/6.

*Isotherm measurements*

Pressure and resistivity isotherms were simultaneously measured by optical transmission $I$ at 625 nm and 4-probe-resistance measurements $\rho$ between 100-225°C in a UHV chamber, described in detail in [27]. Relative changes in optical transmission are derived via the Lambert-Beer Law as -ln($I/I_0$) [50], and the excess resistivity is quantified by ($\rho-\rho_0$)/$\rho_0$ = $\Delta\rho/\rho_0$, where $I_0$ and $\rho_0$ are measurements at UHV conditions. Changes in the optical transmission due to phase transition in the capping layer (Pd) are not expected at these pressures and temperatures [51] [52]. The hydrogen pressure $p$ is increased stepwise, and data is extracted after thermodynamic equilibrium is reached. The calibration of -ln($I/I_0$) to absolute hydrogen done by $^{15}$N-NRA in [28] [29] [30] is of linear correlation, however with small axes offsets in opposite directions for the two superlattice systems. The negative offset for Cr/V has been attributed to a systematic uncertainty arising from a minor drift in optical intensity over prolonged measurement periods and can be omitted in the analysis. The positive offset in Fe/V is attributed to hydrogen first occupying defect sites [30] the presence of which is revealed by the resistivity measurements shown below in Fig. 1 (c) of this work. As the occupation of these lattice defects is not observed to induce a response in -ln($I/I_0$), only the slopes of the calibration curves are used to calibrate the optical transmission. At 625 nm -ln($I/I_0$) converts average hydrogen concentration in vanadium $c$ [H/V] via (1.20±0.06) and (1.06±0.03) for Cr/V and Fe/V, respectively.

*Direct detection of hydrogen concentration and site location*

To investigate hydrogen distribution, site location, and vibrational motion in the superlattices at different hydrogen concentrations, the samples were further prepared for ex-situ ion beam analysis. A thin layer of $Al_2O_3$ was deposited by reactive magnetron sputtering, which slows hydrogen ab- and desorption during the loading procedure and minimizes hydrogen desorption under ion beam exposure [53]. The coated superlattices were loaded with hydrogen, thereby heated to 140°C, and exposed to pressures up to 1 bar of hydrogen gas. After several hours, the chamber was gradually evacuated and cooled to room temperature and samples were subsequently transferred for ion beam analysis. For more details on the procedure, we refer to [29] [54].

Ion beam experiments using RBS and NRA were performed at the 5 MV 15 SDH-2 Tandem accelerator, Uppsala University, Sweden [55] employing a $^{15}N$-ion beam within an energy range of 6.37-6.7 MeV with fluences below 2 nA. Backscattered $^{15}N$-ions are captured by a solid-state detector at a scattering angle of $\varphi = 160°$. The γ(4.43 MeV) radiation resulting from the $^{1}H(^{15}N, \alpha\gamma)^{12}C$ nuclear reaction ($E_R$ = 6.385 MeV) is detected with a Bismuth Germanate scintillation crystal placed outside the experimental chamber behind the sample. Accurate hydrogen concentrations are extracted from NRA excitation profiles positioning the resonance energy at varying depths in the target for non-channeling geometries [25]. As the depth resolution is several nm, hydrogen distribution is averaged over several bilayers of a superlattice. The current normalized γ-yield is calibrated to a H-implanted Si-standard of 18.5%, considering changes in the electronic energy loss cross-section of the metal hydride compound according to the analysis procedure described in [56]. The location of interstitial hydrogen in the unit cells is determined by performing simultaneous channeling $^{15}N$-NRA and $^{15}N$-RBS measurements and Monte-Carlo simulations evaluating multiple crystal planes in a single angular-resolved scan [26]. The analysis of channeling $^{15}N$-RBS through the superlattices, which is presented in Ref. [30], gives a quantitative measure of ion trajectories by fitting ion flux distributions simulated with the Monte Carlo method-based program FLUX7 [57]. Thereby, a normalized average nuclear encounter probability (NEP) of ion projectiles to lattice atoms is derived, which is evaluated against normalized experimental RBS channeling yields. The supporting program YIMP7 [57] calculates the NEP of the $^{15}N$-ion flux with thermally vibrating interstitial hydrogen occupying certain interstitial sites, employed to analyze channeling $^{15}N$-NRA.

**Acknowledgments**

The authors gratefully acknowledge financial support from the Tandem accelerator infrastructure by VR-RFI (contract #2017-00646_9 and #2019_00191) as well as the Swedish Foundation for Strategic Research (SSF) under contract RIF14-0053.